\documentclass[aps,pra,notitlepage,superscriptaddress]{revtex4-1}

\usepackage{epstopdf}
\usepackage[caption=false]{subfig}

\usepackage{wrapfig}
\usepackage{hyperref}
\usepackage{array} 
\usepackage{listings}
\usepackage[para,online,flushleft]{threeparttablex}
\usepackage{booktabs,dcolumn}
\usepackage{textpos}
\usepackage{booktabs}
\usepackage{multirow,bigdelim}
\usepackage{float}
\usepackage{todonotes}
\usepackage{amsfonts}
\usepackage{amsmath}
\usepackage{amssymb}
\usepackage{graphicx}
\usepackage{epstopdf}
\usepackage{bm}
\usepackage{color}
\usepackage{dcolumn}

\renewcommand{\phi}{\varphi}
\newcommand\+{\;\lower\plusheight\hbox{$+$}\;}
\newcommand\lldots{\;\lower\plusheight\hbox{$\cdots$}\;}
\newcommand{\ud}{\mathrm{d}}

\begin{document}

\title{The Madelung Constant in $N$ Dimensions}

\author{Antony Burrows}
\affiliation{Centre for Theoretical Chemistry and Physics, The New Zealand Institute for Advanced Study, Massey University, 0632 Auckland, New Zealand}
\author{Shaun Cooper}
\email[email:]{S.Cooper@massey.ac.nz}
\affiliation{School of Mathematical and Computational Sciences, Massey University, 0632 Auckland, New Zealand.}
\author{Peter Schwerdtfeger}
\email[email:]{peter.schwerdtfeger@gmail.com}
\affiliation{Centre for Theoretical Chemistry and Physics, The New Zealand Institute for Advanced Study, Massey University, 0632 Auckland, New Zealand}

\date{\today}

\begin{abstract}
We introduce two convergent series expansions (direct and recursive) in terms of Bessel functions and representations of sums $r_N(m)$ of squares for $N$-dimensional Madelung constants, $M_N(s)$, where $s$ is the exponent of the Madelung series (usually chosen as $s=1/2$). The functional behavior including analytical continuation, and the convergence of the Bessel function expansion is discussed in detail. Recursive definitions are used to evaluate $r_N(m)$. Values for $M_N(s)$ for $s=\tfrac{1}{2}, \tfrac{3}{2}, 3$ and 6 for dimension up to $N=20$ and for $M_N(1/2)$ up to $N=100$ are presented. Zucker's original analysis [J. Phys. A {\bf 7}, 1568 (1974)] on $N$-dimensional Madelung constants for even dimensions up to $N=8$ and their possible continuation into higher dimensions is briefly analyzed. 
\end{abstract}

\maketitle

\section{Introduction}
\label{Introduction}

The classical lattice energy $E_{\rm lat}$ of an ionic crystal M$^+$X$^-$can be obtained from lattice summations of Coulomb interacting point charges and is usually presented by the Born-Lande form \cite{BornLande1918,BornLande1918a}
\begin{equation}
\label{Born-Lande}
E_{\rm lat}=-\frac{N_AZ^2e^2}{4\pi\epsilon_0R_0}M_{\rm lat}\left( 1-n^{-1}\right) ,
\end{equation}
where $M_{\rm lat}$ is the Madelung constant for a specific lattice \cite{madelung1918}, $N_A$ is Avogadro's constant, and $n$ is the Born exponent which corrects for the repulsion energy $V=aR^{-n}, a>0$ at nearest neighbor distance $R_0$, $Z$ is the ionic charge (+1 in the ideal case), $e$ and $\epsilon_0$ are the elementary charge and vacuum permittivity respectively. Values for $Z^2M$ and $n$ have been tabulated for different crystals in the past \cite{quane1970}. For a simple cubic lattice with alternating charges in the crystal the Madelung constant (or function) $M(s)\equiv M_{\rm sc}(s)$ is given by the 3D alternating lattice sum
\begin{equation}
\label{Madelung}
M(s) = {\sum_{i,j,k\in\mathbb{Z}}}^{\hspace{-.2cm}'}  \hspace{.2cm}\frac{(-1)^{i+j +k}} {(i^2+j^2+ k^2)^s} \quad ,
\end{equation}
where the summation is over all integer values, the prime behind the sum indicates that $i=j=k=0$ is omitted, $s\in\mathbb{R}$, and $s=\frac{1}{2}$ is chosen for a Coulomb-type of interaction. This sum is absolutely convergent for $s>\frac{3}{2}$, but only conditionally convergent for smaller $s$-values \cite{Emersleben1950,borwein1998convergence}. The problem with conditionally convergent series is that the Riemann Series Theorem states that one can converge to any desired value or even diverge by a suitable rearrangement of the terms in the series. This problem is well known for the Madelung constant ($s=\frac{1}{2}$) and has been documented and analyzed in great detail by Borwein et al \cite{borwein1985,borwein1998convergence,borwein-2013} and Crandall et al \cite{Crandall_1987,Crandall1999}. For example, one has to sum over expanding cubes and not spheres to arrive at the correct result of $M(\frac{1}{2})=-1.747~564~594~633~182~\ldots$ \cite{Crandall1999}.

It is currently not known if the Madelung constant can be expressed in terms of standard functions. The closest formula one can get is the one for $s=\frac{1}{2}$ recently derived by Tyagi \cite{Tyagi2005} following an approach by Crandall \cite{Crandall1999},
\begin{align}
\label{Tyagi}
M\left(\tfrac{1}{2}\right) = -\frac{1}{8} - \frac{\ln{2}}{4\pi} - \frac{4\pi}{3} +  \frac{1}{2\sqrt{2}} +  \frac{\Gamma\left(\frac{1}{8}\right)\Gamma\left(\frac{3}{8}\right)}{\pi^{3/2}\sqrt{2}} 
- 2{\sum_{k\in\mathbb{N}}} \: \frac{(-1)^k r_3(k)} {\sqrt{k} \left[ e^{8\pi\sqrt{k}}-1\right]} 
\end{align}
which is correct to 10 significant figures if the sum is neglected (for more recent work and improvement of Tyagi's formula see Zucker \cite{Zucker2013}). Moreover, the sum converges relatively fast. Here $r_3(k)$ is the number of representations of $k$ as a sum of three squares.

There are many expansions available leading to an accurate determination of the Madelung constant  \cite{Crandall1999}. Perhaps the most prominent formulas are the ones by Benson-Mackenzie \cite{benson1956,mackenzie1957}
\begin{equation}
\label{Madelung2}
M\left(\tfrac{1}{2}\right) = -12\pi {\sum_{i,j\in\mathbb{N}}} {\rm sech}^2 \left[ \frac{\pi}{2}\sqrt{(2i-1)^2+(2j-1)^2} \right]
\end{equation}
and by Hautot \cite{Hautot-1975} (in modified form by Crandall  \cite{Crandall1999})
\begin{align}
\label{Madelung3}
M\left(\tfrac{1}{2}\right) &= -\frac{\pi}{2} + 3 {\sum_{i,j\in\mathbb{Z}}}^{'}   \hspace{.4cm} \frac{(-1)^i {\rm cosech} \left( \pi \sqrt{i^2+j^2} \right)} {\sqrt{i^2+j^2}} 
\end{align}

The Madelung constant can easily be extended to a $N$ dimensional series ($N>$0),
\begin{equation}
\label{MadelungN}
M_N(s) =  {\sum_{i_1,\ldots,i_N\in\mathbb{Z}}}^{\hspace{-.4cm}'}  \hspace{.4cm}  \frac{(-1)^{i_1+\cdots + i_N}}{(i_1^2+i_2^2+\cdots + i_N^2)^s }
= {\sum_{\hspace{.1cm} \vec{i}\in\mathbb{Z}^N\backslash \{\vec{0}\}}} \: \frac{(-1)^{\vec{i}\cdot\vec{1}}}{|\vec{i}|^{2s}}
\quad ,
\end{equation}
and the prime after the sum denotes that the term corresponding to $i_1= i_2= \cdots =i_N=0$ is omitted (in the shorter notation on the right $\vec{1}=(1,1,\ldots,1)^\top$ ). The sum is absolutely convergent for exponents $s>\frac{N}{2}$. The Madelung series is as a special case of the more general Epstein zeta function \cite{Epstein-1903}. 

Zucker has found analytical expressions in terms of standard functions for even dimensions up to $N=8$ \cite{Zucker-1974},
\begin{align}
\label{MadelungN1}
M_1(s) &=  -2\eta(2s) \\ 
\label{MadelungN2}
M_2(s) &=  -4\beta(s)\eta(s)\\
\label{MadelungN4}
 M_4(s) &= -8\eta(s-1)\eta(s) \\
 \label{MadelungN6}
M_6(s) &= -16\eta(s-2)\beta(s) + 4\eta(s)\beta(s-2)  \\
\label{MadelungN8}
M_8(s) &= -16\eta(s-3)\zeta(s)
\end{align}
Here $\eta(s)$ is the Dirichlet eta function, $\beta(s)$ the Dirichlet beta function, and $\zeta(s)$ the Riemann zeta function \cite{Zucker-1974}. These standard functions are defined in the Appendix A together with their analytical continuations to the whole range of real (or complex) numbers, $s\in\mathbb{R}(\mathbb{C})$. 

\begin{figure}[ht]
  \begin{center}
 \includegraphics[scale=.35]{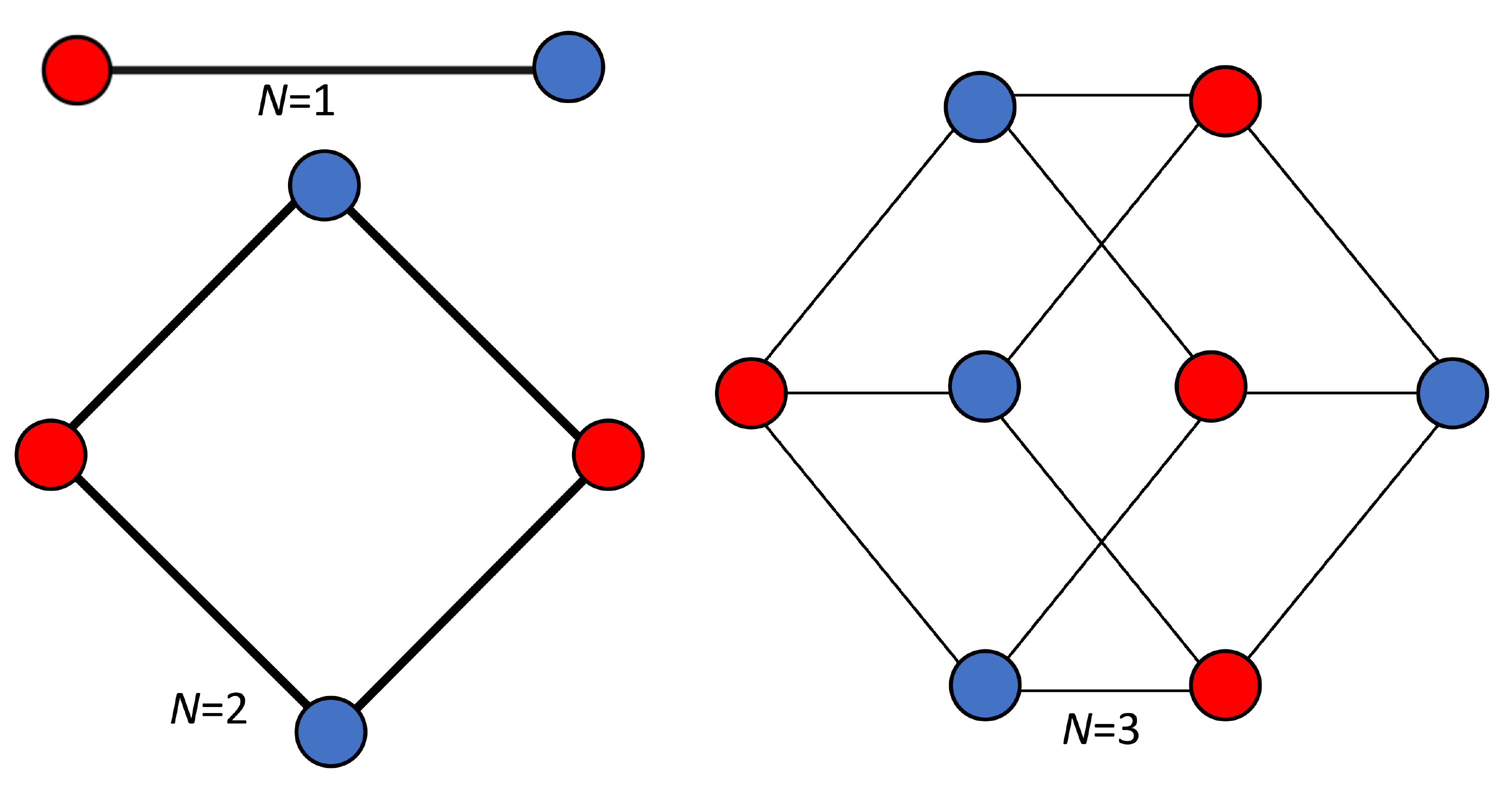}\\
 \includegraphics[scale=.35]{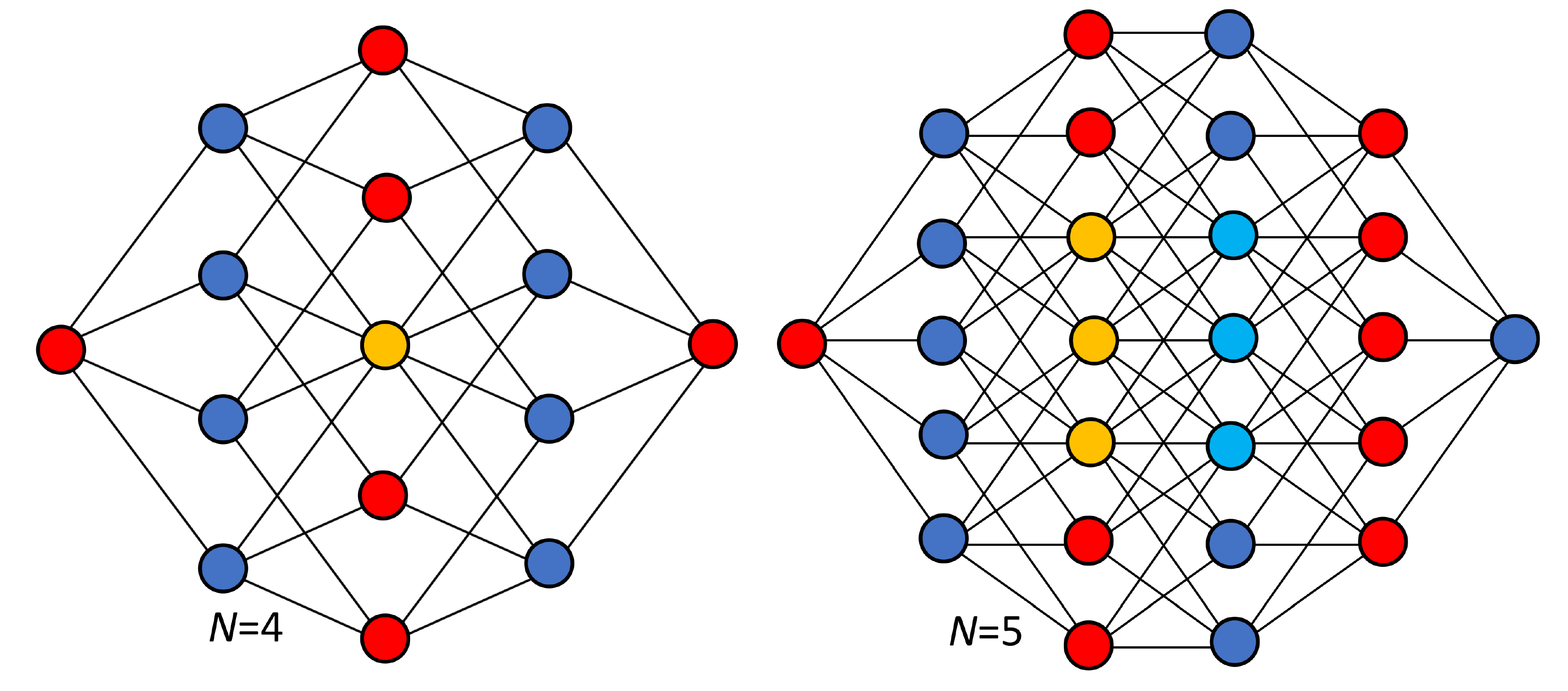}
  \caption{Graphs derived from orthogonal 2D projections of $N$-cubes ($1\le N\le 5$) showing the alternating colors for the vertices ($\pm1$ charges for the atoms). Starting with the 4-cube (tesseract) the orthogonal projection shows vertices overlapping and we use lighter colors to highlight the two overlapping vertices (orange for two red vertices and light blue for the two blue vertices).}
  \label{fig:Ncubes}
  \end{center}
  \end{figure}

By analogy with the three-dimensional case, an $N$-dimensional lattice can easily be constructed from its $N$ linearly independent basis lattice vectors (or transformations of it). Higher dimensional lattices and their properties have been catalogued (up to certain dimensions) by Nebe and Sloane \cite{nebe2012catalogue}. The simple cubic $N$-dimensional lattice can be drawn as an infinite graph with atoms (vertices) and edges connecting the nearest neighbor atoms (adjacent vertices). If we walk around the edges we alternate the charges (+/- sign or red/blue color of the vertices in the graph) in the ionic lattice corresponding to the alternating series for the Madelung constant. We can also derive the lattice from tiling the $N$-dimensional space with $N$-cubes by implying translational symmetry. Figure \ref{fig:Ncubes} shows the graphs for such $N$-cubes up to $N=5$ together with the alternating color scheme. We notice that for dimensions $N>3$ the graphs are not planar anymore. The number of nearest neighbor vertices for an N-dimensional cubic lattice is $2N$ and corresponds to the limit,
\begin{equation}
\label{MadelungLimit}
\lim_{s \to \infty} M_N(s) =  -2N \quad .
\end{equation}
For example, Crandall reports $M_3(50)=-5.999~999~999~999~989~341\ldots$ \cite{Crandall1999}.
  
A general and relatively fast converging series expansion for the $N$-dimensional Madelung constant has been elusive for a very long time. For example, a recent suggestion was made by Mamode to use the Hadamard finite part of the integral representation of the underlying potential (e.g. a Coulomb potential) within the $N$-dimensional crystal \cite{mamode2017}, but computations are quite involved and results presented  were only up to three dimensions. For the $N$-dimensional case one can explore expansions known for example for the Epstein zeta function \cite{terras-1973,Crandall_1987,crandall1998fast} or similar techniques \cite{burrows-2020}. In this work, we introduce a general formula for the $N$-dimensional Madelung constant for a simple cubic crystal in terms of a fast convergent Bessel function expansion allowing for analytical continuation, which gives deep insight into the functional behavior of the $N$-dimensional Madelung constant. The derivation is given in the next section. The convergence of $M_N(s)$ with increasing dimension $N$ is discussed in detail in the results section.

\section{Theory}
\label{Theory}

In this section we derive two useful expansions for the $N$-dimensional Madelung constant. 
Consider $M_{N+1}(s) $ and change the last summation index to $k$, and write
\begin{equation}
\label{ne1}
M_{N+1}(s) 
= {\sum_
{\substack{i_1,\ldots,i_N\in \mathbb{Z} \\ k\in \mathbb{Z}}}}
^{\!\!\!\!\!\!\!\prime} \;\;\;\; \frac{(-1)^{i_1+\cdots + i_N+k}}{(i_1^2+i_2^2+\cdots + i_N^2
+k^2)^s}.
\end{equation}
Now separate the sum into the two cases
$k=0$ and $k\neq 0$ to get
\begin{equation}
\label{ne1a}
M_{N+1}(s)  = M_{N}(s) +2F(s)
\end{equation}
where
\begin{equation}
\label{ne2}
F(s) = \sum_{k\in\mathbb{N}} \left(
\sum_{i_1,\ldots,i_N\in \mathbb{Z}} \; \frac{(-1)^{i_1+\cdots + i_N+k}}{(i_1^2+i_2^2+\cdots + i_N^2+k^2)^s}\right).
\end{equation}
By the gamma function integral in the form ($\mathbb{R}_+ = \{ x\in\mathbb{R}~ |~ x\ge0 \}$) 
\begin{equation}
\label{ne3}
\frac{1}{z^s} = \frac{1}{\Gamma(s)} \int_{\mathbb{R}_+} t^{s-1}e^{-zt}\,\ud t
\end{equation}
we have
\begin{align}
\nonumber
\pi^{-s}&\Gamma(s)F(s)=
 \int_{\mathbb{R}_+} t^{s-1} \left(\sum_{k\in \mathbb{N}} (-1)^ke^{-\pi k^2 t}\right)
\left(\sum_{i_1,\ldots,i_N\in \mathbb{Z}} (-1)^{i_1+\cdots+i_N}e^{-\pi (i_1^2+\cdots + i_N^2)t}\right) \ud t \\
=& \int_{\mathbb{R}_+} t^{s-1} \left(\sum_{k \in \mathbb{N}} (-1)^ke^{-\pi k^2 t}\right)
\left(\sum_{j\in \mathbb{Z}} (-1)^je^{-\pi j^2t}\right)^N \;\ud t.
\end{align}
By using the modular transformation for the theta function \cite{andrews1999special}, 
\begin{equation}
\sum_{n\in \mathbb{Z}} e^{-\pi n^2t+2\pi ina} = \frac{1}{\sqrt{t}} \sum_{n\in \mathbb{Z}} e^{-\pi (n+a)^2/t} 
\end{equation}
we get
\begin{align}
\pi^{-s}&\Gamma(s)F(s)=
 \int_{\mathbb{R}_+} t^{s-1} \left(\sum_{k\in \mathbb{N}} (-1)^ke^{-\pi k^2 t}\right)
\left(\frac{1}{\sqrt{t}}\,\sum_{j\in \mathbb{Z}} e^{-\pi (j+\frac12)^2/t}\right)^N \;\ud t.
\end{align}
This can be rearranged further to give
\begin{align}
\label{ny20}
\pi^{-s}&\Gamma(s)F(s) =
 \int_{\mathbb{R}_+} t^{s-1-\frac{N}{2}} \left(\sum_{k\in \mathbb{N}} (-1)^ke^{-\pi k^2 t}\right)
\left(\sum_{\:\: m\in \mathbb{N}_0} r_{N}^{\text{odd}}(8m+N)\,e^{-\pi (8m+N)/4t}\right) \;\ud t
\end{align}
where $\mathbb{N}_0$ denotes the natural numbers including zero, and $r_{N}^{\text{odd}}(m)$ is the number of representations of $m$ as a sum of~$N$ odd squares. That is, $r_{N}^{\text{odd}}(m)$ is the number of solutions of
\begin{equation}
(2j_1+1)^2+(2j_2+1)^2+\cdots +(2j_N+1)^2 = m
\end{equation}
in integers. 
The integral in (\ref{ny20}) can be evaluated in terms of Bessel functions by means of the formula
\begin{equation}
\label{BesselK}
\int_{\mathbb{R}_+} t^{\nu-1}e^{-at-b/t} \ud t = 2\left(\frac{b}{a}\right)^{\nu/2} K_\nu(2\sqrt{ab}).
\end{equation}
to give
\begin{align}
\pi^{-s}&\Gamma(s)F(s)=
2\sum_{k\in \mathbb{N}}  \sum_{\;\; m\in \mathbb{N}_0}  (-1)^k r_{N}^{\text{odd}}(8m+N)\,
\left(\frac{8m+N}{4k^2}\right)^{(2s-N)/4} \, K_{s-N/2}\left(\pi k \sqrt{8m+N}\right)~.
\end{align}
On using this result back in~\eqref{ne1} we obtain the recursion relation for the Madelung constant in terms of the dimension $N$,
\begin{align}
\label{Madelungrec}
 &M_{N+1}(s)  = M_{N}(s)
  + \frac{4 \pi^s}{\Gamma(s)} \sum_{k\in \mathbb{N}}  \sum_{\:\: m\in \mathbb{N}_0}  (-1)^k r_{N}^{\text{odd}}(8m+N)\,
\left(\frac{8m+N}{4k^2}\right)^{(2s-N)/4} \, K_{s-N/2}\left(\pi k \sqrt{8m+N}\right)\\
 \nonumber
& = M_{N}(s) +  \sum_{\:\: m\in \mathbb{N}_0}  r_{N}^{\text{odd}}(8m+N) c_{s,N}(m)
\end{align}
with
\begin{align}
\label{Madelungrec1}
c_{s,N}(m) = \frac{4 \pi^s}{\Gamma(s)} \sum_{k\in \mathbb{N}}   (-1)^k \left(\frac{8m+N}{4k^2}\right)^{(2s-N)/4} \, K_{s-N/2}\left(\pi k \sqrt{8m+N}\right)~ .
\end{align}
For fixed $N$, the term $r_{N}^{\text{odd}}(8m+N)$ can become very large for larger $m$ and $N$ values, but is more than compensated by the exponentially decreasing Bessel function, which we discuss in detail in the next section. The $r_N^{\text{odd}}(m)$ values can be determined recursively which is described in the Appendix.

While the recursion relation (\ref{Madelungrec}) is useful if the Madelung constant of lower dimension is known, we seek for a second formula where the recursion relation has been resolved. Here, we proceed as above and separate the sum for $M_{N+1}(s)$ into two cases according to whether $i_1=i_2=\cdots=i_N=0$ or $i_1$, $i_2$, $\ldots$, $i_N$ are not all zero. This gives
\begin{equation}
\label{e31}
M_{N+1}(s)  =2\sum_{k\in \mathbb{N}} \frac{(-1)^k}{k^{2s}}+g(s)
\end{equation}
where 
\begin{equation}
\nonumber
g(s) = \sum_{k\in \mathbb{Z}} \left( {\sum_{i_1,\ldots,i_N\in \mathbb{Z}}}^{\!\!\!\!\!\prime} \;\;\: 
\frac{(-1)^{i_1+\cdots + i_N+k}}{(i_1^2+i_2^2+\cdots + i_N^2+k^2)^s} \right).
\end{equation}
Applying the integral formula for the gamma function and then the modular transformation for the theta function we obtain
\begin{align}
\pi^{-s}\Gamma(s)g(s)
&= \int_{\mathbb{R}_+} t^{s-1} {\sum_{i_1,\ldots,i_N\in \mathbb{Z}}}^{\!\!\!\!\!\prime} \;\;
(-1)^{i_1+\cdots + i_N}e^{-\pi(i_1^2+\cdots +i_N^2)t}
\sum_{k\in \mathbb{Z}} (-1)^ke^{-\pi k^2 t} \, \ud t \\
\nonumber
&= \int_{\mathbb{R}_+} t^{s-3/2} {\sum_{i_1,\ldots,i_N\in \mathbb{Z}}}^{\!\!\!\!\!\prime} \;\;
(-1)^{i_1+\cdots + i_N}e^{-\pi(i_1^2+\cdots +i_N^2)t}
\sum_{k\in \mathbb{Z}} e^{-\pi (k+\frac12)^2/t} \, \ud t \\
\nonumber
&= 2\int_{\mathbb{R}_+} t^{s-3/2} {\sum_{i_1,\ldots,i_N\in \mathbb{Z}}}^{\!\!\!\!\!\prime} \;\;
(-1)^{i_1+\cdots + i_N}e^{-\pi(i_1^2+\cdots +i_N^2)t}
\sum_{k\in \mathbb{N}} e^{-\pi (k-\frac12)^2/t} \, \ud t,
\end{align}
where the last step follows by noting
\begin{equation}
\sum_{k\in \mathbb{Z}} e^{-\pi (k+\frac12)^2/t} = 2\sum_{k\in \mathbb{N}_0} e^{-\pi (k+\frac12)^2/t} = 2\sum_{k \in \mathbb{N}} e^{-\pi (k-\frac12)^2/t}.
\end{equation}
In terms of the modified Bessel function this becomes, by~\eqref{BesselK},
\begin{align}
&\pi^{-s}\Gamma(s)g(s)=
4 {\sum_{i_1,\ldots,i_N\in \mathbb{Z}}}^{\!\!\!\!\!\prime} \;\;\; \sum_{k\in \mathbb{N}}
(-1)^{i_1+\cdots + i_N}
\left(\frac{k-\frac12}{\sqrt{i_1^2+\cdots+i_N^2}}\right)^{s-\frac12}
K_{s-\frac12}\left(2\pi (k-\frac12)\sqrt{i_1^2+\cdots +i_N^2}\right) \\
\nonumber
&= 4\sum_{m\in \mathbb{N}} \sum_{k\in \mathbb{N}} (-1)^m r_N(m) 
\left(\frac{k-\frac12}{\sqrt{m}}\right)^{s-\frac12}
K_{s-\frac12}\left(2\pi (k-\frac12)\sqrt{m}\right).
\end{align}
On using this back in~\eqref{e31} we obtain
\begin{align}
\label{e32}
&M_{N+1}(s)  =
-2\eta(2s) + \frac{4\pi^s}{\Gamma(s)} \sum_{m\in \mathbb{N}}  (-1)^m r_N(m) \sum_{k\in \mathbb{N}} 
\left(\frac{k-\frac12}{\sqrt{m}}\right)^{s-\frac12} K_{s-\frac12}\left(\pi (2k-1)\sqrt{m}\right).
\end{align}
For the case of $N=0$ the sum of the right-hand side is zero ($r_0(m)=0$) and we have $M_1(s)  =-2\eta(2s)$ in agreement with Zucker's formula (\ref{MadelungN1}). We can conveniently write the sum in the form,
\begin{equation}
\label{e33}
M_{N+1}(s)  = -2\eta(2s) + \sum_{m\in \mathbb{N}} (-1)^m r_N(m) c_s(m)
\end{equation}
with
\begin{equation}
\label{e34}
c_s(m)=\frac{4\pi^s}{\Gamma(s)} m^{\frac{1-2s}{4}} \sum_{k\in \mathbb{N}} 
\left( k-\frac12 \right)^{s-\frac12} K_{s-\frac12}\left(\pi (2k-1)\sqrt{m}\right)
\end{equation}
Note that the coefficients $c_s(m)$ are independent of the dimension $N$. The sum in (\ref{e34}) converges fast because of the exponential asymptotic decay of the Bessel function. The more problematic part is the convergence with respect to the first sum (see eq.\ref{e33}) over $m$ as we shall see. 

As a special case we evaluate $M_N(1/2)$. Letting $s\rightarrow 1/2$ in~\eqref{e32} gives a formula for the $N+1$ dimensional Madelung constant
\begin{equation}
\label{N-Madelung}
M_{N+1}(1/2)  = -2 \ln 2 + 4\sum_{m\in \mathbb{N}} \sum_{\; k\in \mathbb{N}} (-1)^m r_N(m) K_{0}\left(\pi (2k-1)\sqrt{m}\right)
\end{equation}
where $r_N(m)$ is the number of representations of $m$ as a sum of $N$ squares. The coefficient $c_{1/2}(m)$ becomes
\begin{equation}
\label{N-Madelung2}
c_{1/2}(m) = 4\sum_{k\in \mathbb{N}}
K_{0}\left(\pi (2k-1)\sqrt{m}\right) = 2\int_{\mathbb{R}_+} \frac{1}{\sinh(\pi \sqrt{m} \cosh t)} ~\ud t.
\end{equation}
where the integral is obtained using the formula \cite{temme1996}
\begin{equation}
\label{Bessel}
K_{0}(z) = \int_{\mathbb{R}_+} e^{-z\hspace{.1cm}{\rm cosh}(t)} ~\ud t.
\end{equation}
and summing the resulting geometric series. For example, taking $N=2$ gives
\begin{align}
M_{3}(1/2)  = -2 \ln 2 + 4\sum_{m\in \mathbb{N}} \sum_{\; k\in \mathbb{N}} (-1)^m r_2(m) K_{0}\left(\pi (2k-1)\sqrt{m}\right)
\end{align}
On the other hand, using (\ref{Madelungrec}) and Zucker's equation (\ref{MadelungN1}) we get
\begin{align}
\label{Madelungrecdim3}
 &M_{3}(1/2)  =  -4\beta(1/2)\eta(1/2) 
  + 4 \sum_{k\in \mathbb{N}}  \sum_{\;\; m\in \mathbb{N}_0}  (-1)^k r_{2}^{\text{odd}}(8m+2)\,
\left(\frac{2k^2}{4m+1}\right)^{1/4} \, K_{1/2}\left(\pi k \sqrt{8m+2}\right)
\end{align}

\section{Results}
\label{results}

\begin{figure}[htb]
  \begin{center}
 \includegraphics[scale=.8]{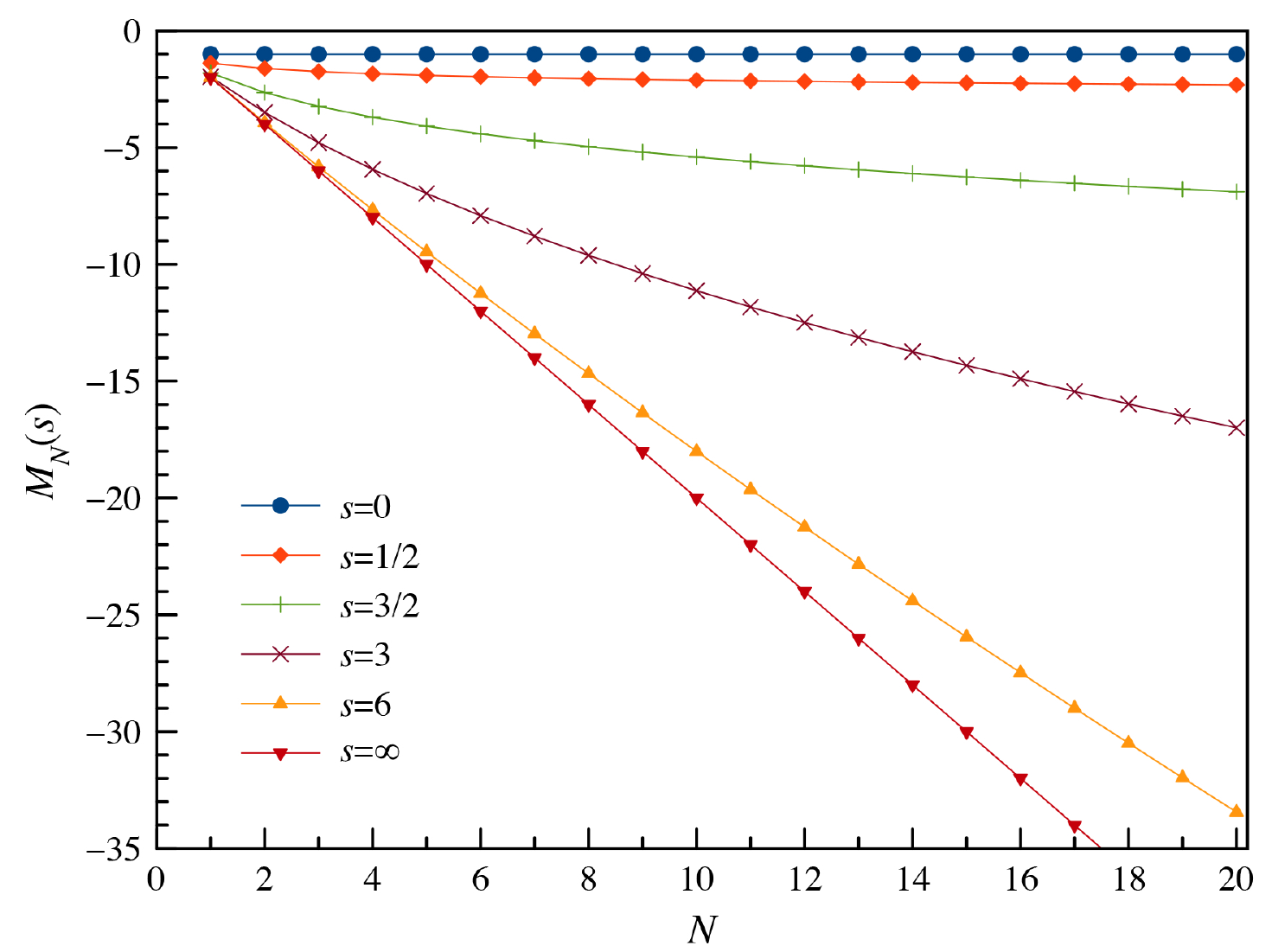}
  \caption{Madelung constants, $M_N(s)$, as a function of the dimension $N$.}
  \label{fig:Madfung}
  \end{center}
  \end{figure}

\begin{figure}[htb]
  \begin{center}
  \includegraphics[scale=.8]{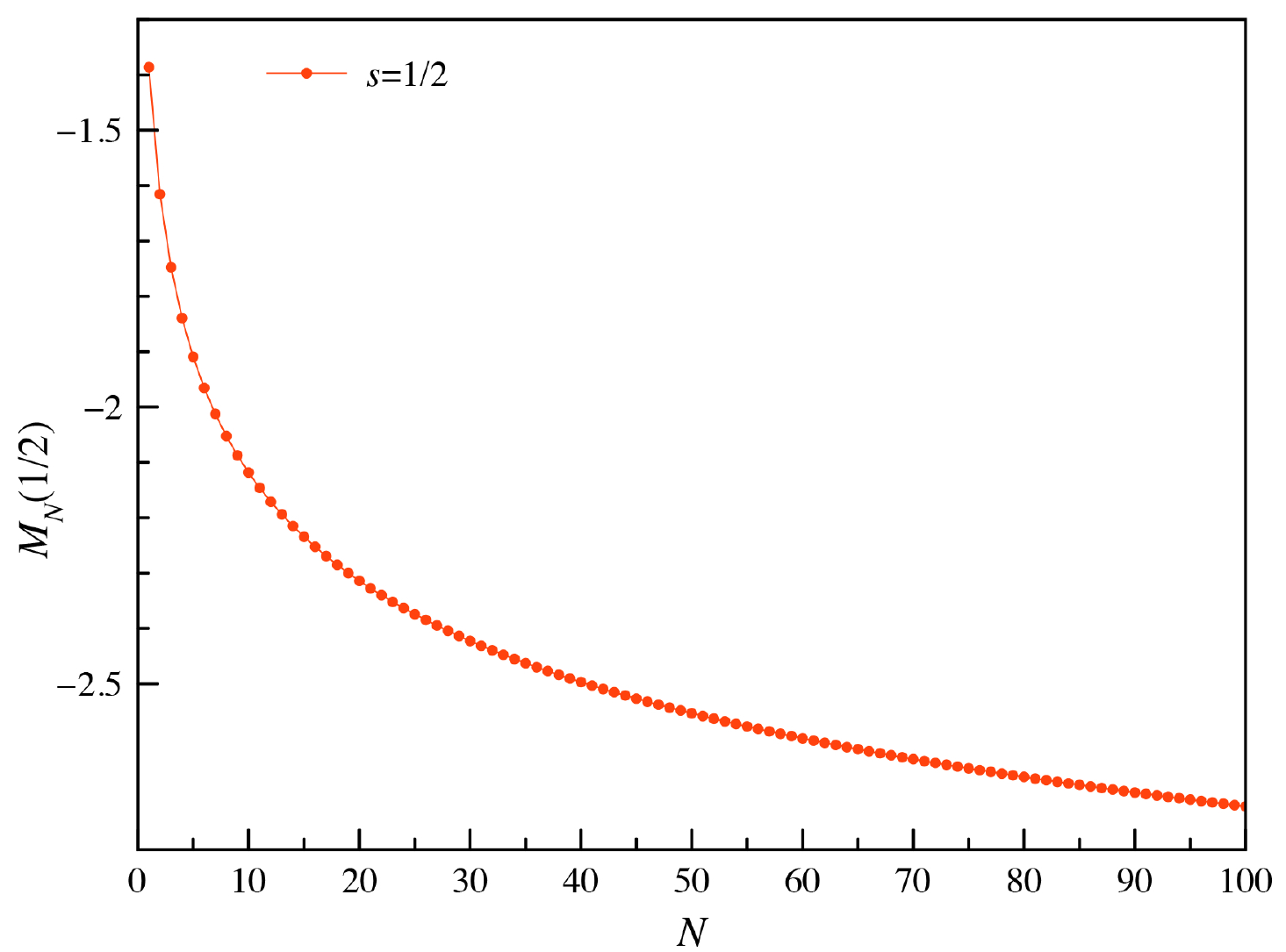}
  \caption{Madelung constants, $M_N(1/2)$, as a function of the dimension $N$ up to $N=100$.}
  \label{fig:Madfunga}
  \end{center}
  \end{figure}

The coefficients $c_{1/2}(m)$ are listed in Table \ref{tab:coeff} together with few selected $r_N(m)$ values. The Madelung constants $M_N(s)$ for selected $s$ values up to dimension $N=20$ are listed in  Table \ref{tab:madelung} and are depicted in Figures \ref{fig:Madfung} and \ref{fig:Madfunga}. The coefficients $c_s(M)$ are all positive for $s>0$, which implies through (\ref{Madelungrec}) that $M_N(s)>M_{N+1}(s)$ for $s>0$. For $N=3$ and $s=1/2$ the Madelung constant is readily evaluated to  computer precision (summing $1\leq m \leq 117$ to reach 14 significant digits (we chose $1 \leq k \leq 200$) as $M_{3}(1/2)   = -1.74756459463318$ in agreement with the known value of Madelung's constant \cite{Crandall1999}. For larger exponents the series converges much faster, i.e. for $M_{3}(6)$ (Table \ref{tab:madelung}) we need to sum only over $1\leq m \leq 51$ to reach convergence to 14 significant digits behind the decimal point. Note that we used backwards summation as small numbers add up. We also checked our values for the even dimensions up to $N=8$ with the values obtained from the analytical function in (\ref{MadelungN1}) by Zucker \cite{Zucker-1974}, and they are in perfect agreement.

\begin{table}
\label{tab:coeff}
\begin{tabular}{ |r|c|r|r|r|r|r|r| } 
 \hline
 $m$ & $c_{1/2}(m)$ & $r_2(m)$ & $r_3(m)$ & $r_4(m)$ & $r_6(m)$ & $r_8(m)$ & $r_{10}(m)$\\ 
 \hline
 1 & 1.18165052269629$\times$10$^{-1}$	& 4 		& 6		& 8 		& 12 		& 16 		& 20 \\
 2 & 2.72719460116136$\times$10$^{-2}$ 	& 4 		& 12		& 24 		& 60 		& 112 	&180 \\
 3 & 9.11805054978030$\times$10$^{-3}$ 	& 0  		&   8		& 32 		& 160 	& 448 	& 960 \\
 4 & 3.66634491506766$\times$10$^{-3}$ 	& 4  		&   6		& 24 		& 252 	& 1136 	& 3380\\
 5 & 1.65469973003050$\times$10$^{-3}$ 	& 8  		&  24		& 48 		& 312 	& 2016 	& 8424 \\
 6 & 8.09716792986126$\times$10$^{-4}$  	& 0  		&  24		& 96 		& 544 	& 3136 	& 16320\\
 7 & 4.21007519555378$\times$10$^{-4}$ 	& 0            & 0         	& 64		& 960	& 5504	& 28800\\
 8 & 2.29579583843101$\times$10$^{-4}$ 	& 4          	& 12        	& 24		& 1020	& 9328	& 52020\\
 9 & 1.30128289377942$\times$10$^{-4}$ 	& 4          	& 30         	& 104	&  876	& 12112	& 88660\\
 10 & 7.61717027007281$\times$10$^{-5}$ 	& 8          	& 24         	& 144	& 1560	& 14112	& 129064\\
 11 & 4.58237287636094$\times$10$^{-5}$ 	& 0          	& 24         	& 96		& 2400	& 21312	& 175680\\
 12 & 2.82249344482993$\times$10$^{-5}$ 	& 0           	& 8		& 96		& 2080	& 31808	& 262080\\
 13 & 1.77472886511553$\times$10$^{-5}$ 	& 8          	& 24         	& 112 	& 2040	& 35168	& 386920\\
 14 & 1.13644088647490$\times$10$^{-5}$ 	& 0          	& 48         	& 192	& 3264	& 38528	& 489600\\
 15 & 7.39644406563549$\times$10$^{-6}$ 	& 0           	& 0         	& 192	& 4160	& 56448	& 600960\\
 16 & 4.88482197748104$\times$10$^{-6}$ 	& 4		& 6		& 24		& 4092	& 74864	& 840500\\
 17 & 3.26906868046647$\times$10$^{-6}$ 	& 8          	& 48         	& 144	& 3480	& 78624	& 1137960\\
 18 & 2.21430457563634$\times$10$^{-6}$ 	& 4          	& 36        	& 312	& 4380	& 84784	& 1330420\\
 19 & 1.51652113308388$\times$10$^{-6}$ 	& 0          	& 24         	& 160	& 7200	& 109760	& 1563840\\
 20 & 1.04924116314272$\times$10$^{-6}$ 	& 8    	&  24    	& 144 	& 6552 	& 143136	& 2050344\\
 40 & 2.62596820286192$\times$10$^{-9}$ 	& 8  		&  24		& 144 	& 26520 	& 1175328 	& 32826664\\
 60 &	 2.73153353546195$\times$10$^{-11}$	& 0           	& 0        	& 576	& 54080	& 4007808	& 164062080\\
 80 & 5.89549945570033$\times$10$^{-13}$	& 8     	&  24  	& 144 	& 106392	& 9432864 	& 525104424 \\
100&	 2.02339226243198$\times$10$^{-14}$	& 12        	& 30        	& 744	& 164052	& 17893136	& 1282320348\\
120& 9.64273816463316$\times$10$^{-16}$ 	& 0     	& 48       	& 576	& 213824	& 32909184	& 2625594240\\
140& 5.88915444967014$\times$10$^{-17}$	& 0           	& 48       	& 1152	& 324480	& 49238784	& 4921862400\\
160& 4.37540432127918$\times$10$^{-18}$	& 8        	& 24      	& 144	& 425880	& 75493152	& 8402122024\\
180& 3.81438178722105$\times$10$^{-19}$ 	& 8           	& 72      	& 1872	& 478296	& 108353952	& 13297454504\\
200& 3.80087523208009$\times$10$^{-20}$ 	& 12      	& 84     	& 744	& 664020 & 146925328	& 20513309148\\
 \hline
\end{tabular}
  \caption{Coefficients $c_{1/2}(m)$ for exponent $s=1/2$ and representations $r_N(m)$ for a number of $m$ and $N$ values.}
\end{table}

\begin{table}
\label{tab:madelung}
\begin{tabular}{ |r|r|r|r|r|r|r| }
 \hline
 $N$ & $m_{\rm max}$&  $M_N(1/2)$ & $M_N(3/2)$ & $M_N(3)$ & $M_N(6)$\\ 
 \hline
 1 	& 0		&  -1.38629436111989 	&  -1.80308535473939 	& -1.97110218259487 	& -1.99951537028771\\ 
 2 	& 101	&  -1.61554262671282 	&  -2.64588653230643	& -3.49418521170288	& -3.93702124248001\\ 
 3 	& 117	&  -1.74756459463318 	&  -3.23862476605177	& -4.78844371389142 	& -5.82302778890550\\ 
 4 	& 135	&  -1.83939908404504 	&  -3.70269117771204	& -5.93191305089188	& -7.66458960508610\\ 
 5 	& 158	&  -1.90933781561876 	&  -4.08665230978501	& -6.96536812867633	& -9.46689838517490\\ 
 6 	& 184	&  -1.96555703900907 	&  -4.41541406455743	& -7.91367677818339	& -11.2339815395894\\ 
 7 	& 212	&  -2.01240598979798 	& - 4.70360905429867	& -8.79344454973204 	& -12.9690759046272\\ 
 8 	& 240	&  -2.05246682726927 	&  -4.96062369646463	& -9.61645522527675	& -14.6748510064791\\ 
 9 	& 268	&  -2.08739431267374 	&  -5.19286448579961	&-10.3914475289766	& -16.3535526240382\\ 
 10	& 302	&  -2.11831050138482 	&  -5.40491155391300	&-11.1251231380028	& -18.0071001619883\\
 11	& 338	&  -2.14601010324383 	&  -5.60015959755479	&-11.8227595210275	& -19.6371554488071\\ 
 12	& 375	&  -2.17107583567180 	&  -5.78119850773166	&-12.4886029215377	& -21.2451729486919\\
 13	& 415	&  -2.19394722663803 	&  -5.95005160868701	&-13.1261312983588	& -22.8324373927323\\
 14	& 458	&  -2.21496368855843 	&  -6.10833126513306	&-13.7382364790321	& -24.4000926119446\\
 15	& 504	&  -2.23439258374969	&  -6.25734417113144	&-14.3273540620924	& -25.9491640475311\\
 16	& 552	&  -2.25244813503955 	&  -6.39816474499813	&-14.8955583649474	& -27.4805766108785\\
 17	& 603	&  -2.26930453765447	&  -6.53168761111553	&-15.4446333073194	& -28.9951690545215\\
 18	& 657	&  -2.28510527781503	&  -6.65866596401893	&-15.9761263123420	& -30.4937056794534\\
 19	& 714	&  -2.29996989965861	&  -6.77974015828765	&-16.4913899618245	& -31.9768859775816\\
 20	& 773	&  -2.31399901326838	&  -6.89545937988985	&-16.9916146519184	& -33.4453526516541\\
 \hline
\end{tabular}
  \caption{Calculated Madelung constants $M_N(s)$ up to dimension $N=$20 for selected $s$ values. The last digit has not been rounded. $m_{\rm max}$ is the maximum integer value in the sum over $m$ in eq.(\ref{e33}), where the remainder $R_{(m_{\rm max}+1)}<10^{-14}$ for $M_N(1/2)$.}
\end{table}

\begin{figure}[htb]
  \begin{center}
 \includegraphics[scale=.8]{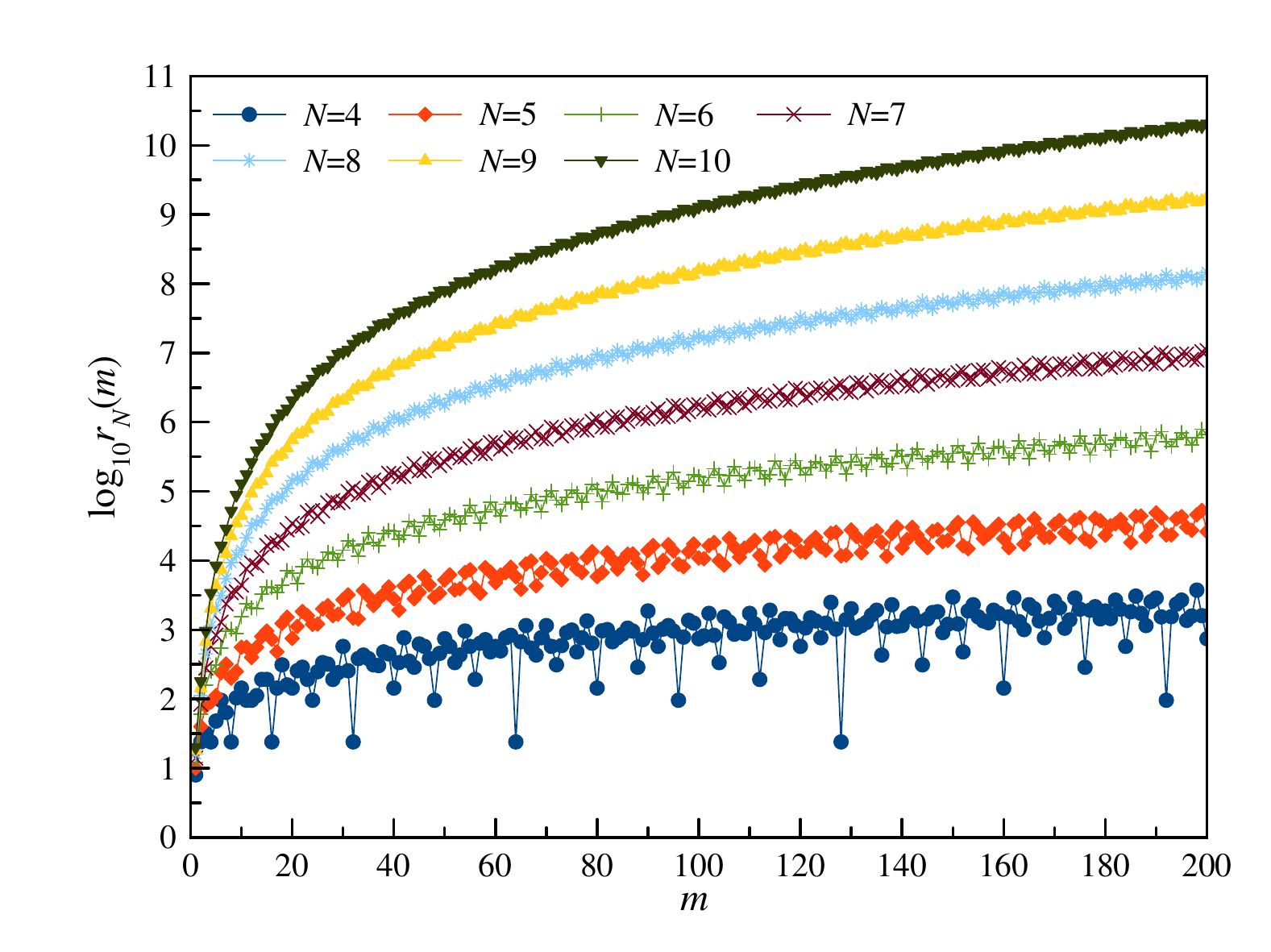}
  \caption{Representations for number of squares, $r_N(m)$, for dimensions $N=4-10$.}
  \label{fig:Madfunc}
  \end{center}
  \end{figure}

  To discuss the convergence behavior of the series (\ref{e33}) we observe that the coefficients $c_{1/2}(m)$ are rapidly decreasing with increasing $m$. However, at the same time the $r_N(m)$ values increase also rapidly with increasing $m$ (and increasing $N$) shown in Figure \ref{fig:Madfunc}. The asymptotic behavior of the Bessel functions is well known, i.e.they decrease exponentially with increasing $m$, $K_s(x)\sim (\pi/2x)^{\tfrac{1}{2}}e^{-x}$. On the other hand, the sum of squares representation increases polynomially for fixed $N$ \cite{hardy1920,rankin1965,holley2019}, e.g. we know from Ramanujan's work that $r_{2N}(m)=\mathcal{O}(m^N)$ (derived from eq.(14) in ref.\cite{ramachandra1987srinivasa}). This is also seen in the logarithmic behavior of ${\rm log}_{10}r_N(m)$ in Figure \ref{fig:Madfunc}. This implies that the Madelung series expansion in terms of Bessel functions is converging, but very slowly for higher dimensions because of a very large dimensional prefactor. This can clearly seen from the $m_{\rm max}$ values for $M_N(1/2)$ in Table \ref{tab:madelung}. For $M_N(s), s\ge 1/2$ we approximately have $m_{\rm max}\le {\rm nint}(1.16N^2+11.5N+73)$, where ${\rm nint}$ represents the nearest integer function. 
  
\begin{figure}[htb]
  \begin{center}
 \includegraphics[scale=.8]{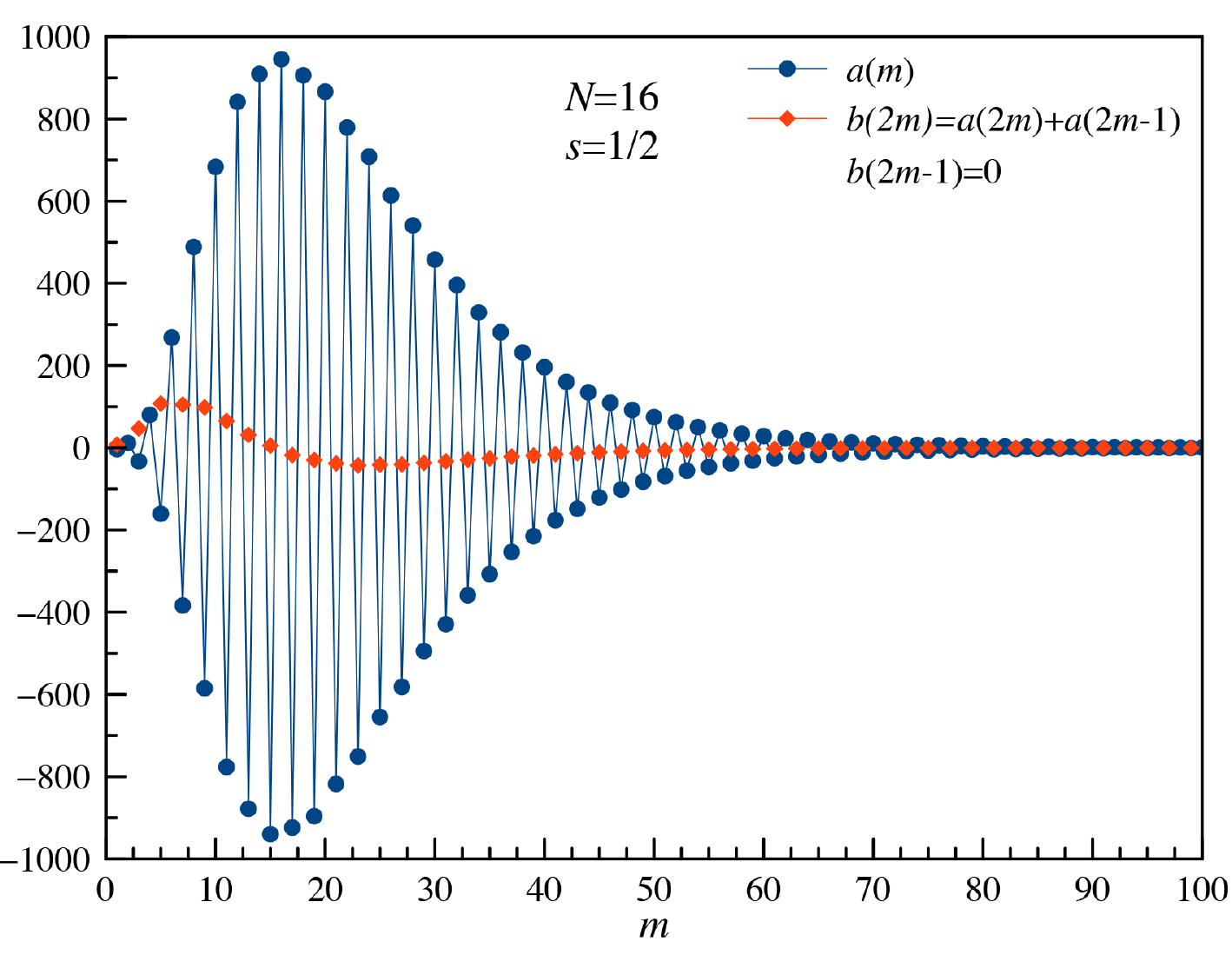}
  \caption{Convergence behaviour for for the Madelung constant with $s=1/2$ and $N=16$. Shown are the coefficients $a(m)=(-1)^m r_{15}(m) c_s(m)$ of eq.(\ref{e33}) (in blue) and the corresponding coefficients by adding the odd and even terms, $b(2m)=a(2m)+a(2m-1)$ (in red). The sum of these values converge against the Bessel sum value of $-0.866153773918593$.}
  \label{fig:convergence}
  \end{center}
  \end{figure}

 Perhaps more problematic is the appearance of large numbers due to the $r_N(m)$ values in the sum over $m$ in eq.(\ref{e33}) where one reaches soon the limit with double precision arithmetic at large $N$ values. This is clearly seen in Figure \ref{fig:convergence} for the case of dimension 16 and $s=1/2$ which shows for the individual terms a strong oscillating behavior and poynomial increase up to rather large values around $m=14$ followed by an exponential decay. For higher dimensions this maximum shifts to higher $m$ values before the exponential decay sets in. However, if we add pairs of positive and negative terms in the oscillating series to obtain new coefficients $b(2m)=a(2m)+a(2m-1)$, we experience a far smoother and better convergence behavior.

\begin{figure}[htb]
  \begin{center}
 \includegraphics[scale=.8]{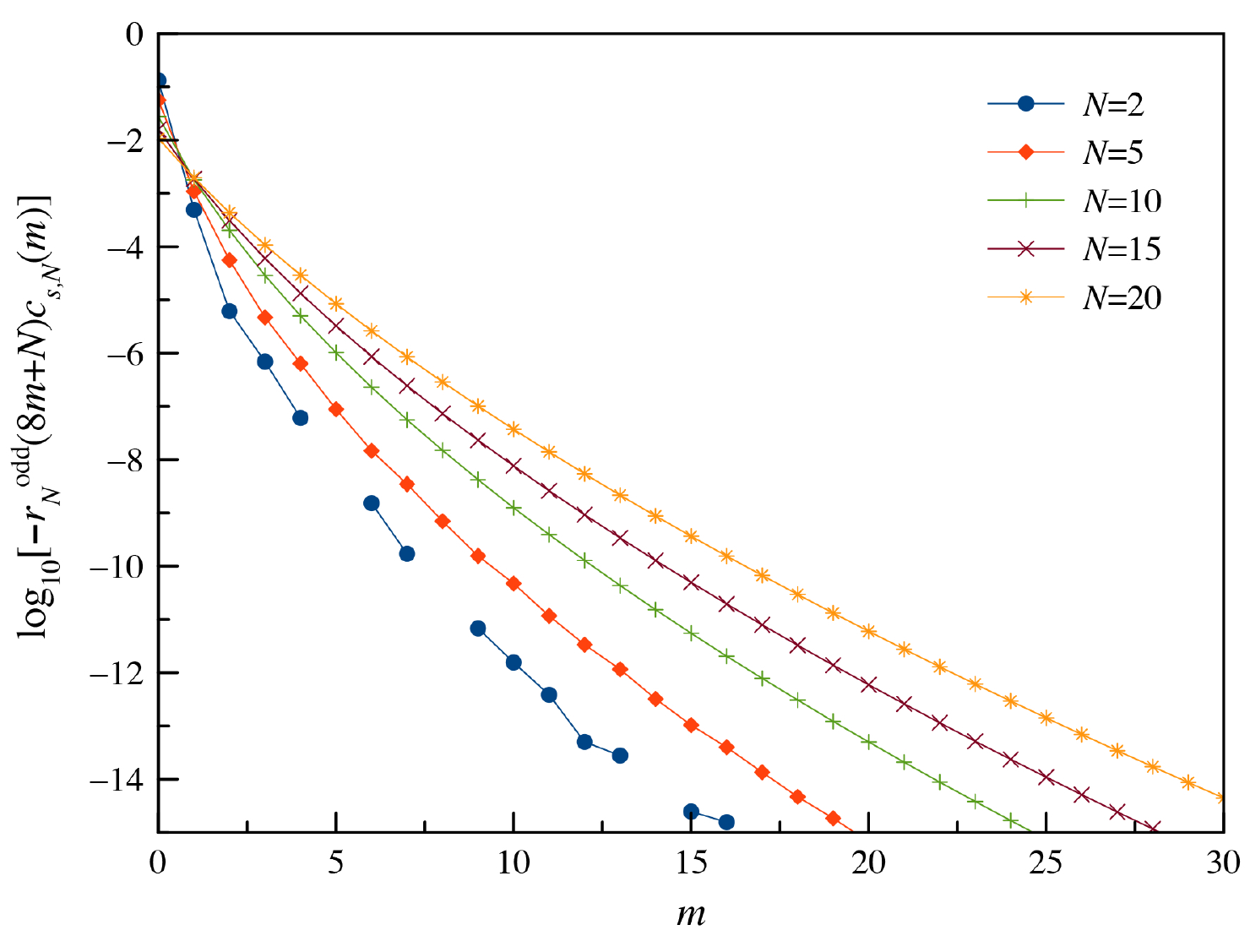}
  \caption{Convergence behaviour for the Madelung constants with $s=1/2$. Shown are the numbers $log_{10}[-d(m)]$ with the coefficients $d(m)= r_{N}^{\rm odd}(8m+N) c_{s,N}(m)$ from eq.(\ref{Madelungrec}). For $N=2$ the missing points have zero value for $r_{N}^{\rm odd}(8m+N)$.}
  \label{fig:convergence1}
  \end{center}
  \end{figure}

By using the recursive formula (\ref{Madelungrec}) instead we obtain much fast convergence as we reach the exponential decay far earlier because of the argument $8m+N$ in the Bessel function, see Figure \ref{fig:convergence1}. Here we avoid such large values and the strong oscillating behavior as the sign change appears in the summation over $k$ in .(\ref{e33}) rather than in (\ref{Madelungrec}). Hence, for accuracy reasons eq.(\ref{Madelungrec}) is preferred, and we used this equation instead for the values in Table \ref{tab:madelung}.

Concerning the analytical continuation all standard functions used including the Bessel function, gamma function and the Dirichlet eta function can be analytically continued (see Appendix) as shown in Figure \ref{fig:Madfunc1}. Moreover, the Madelung constants $M_N(s)$ are all smooth functions without any singularities for all $s\in\mathbb{R}$. For example, from Zucker's formula of $M_8(s) = -16\eta(s-3)\zeta(s)$  we see that for $s=1$ we have $\zeta(1)=\infty$ and $\eta(-2)=0$. However, it can easily be shown that the product of the two functions gives a finite value for $s=1$.
\begin{figure}[htb]
  \begin{center}
\includegraphics[scale=.3]{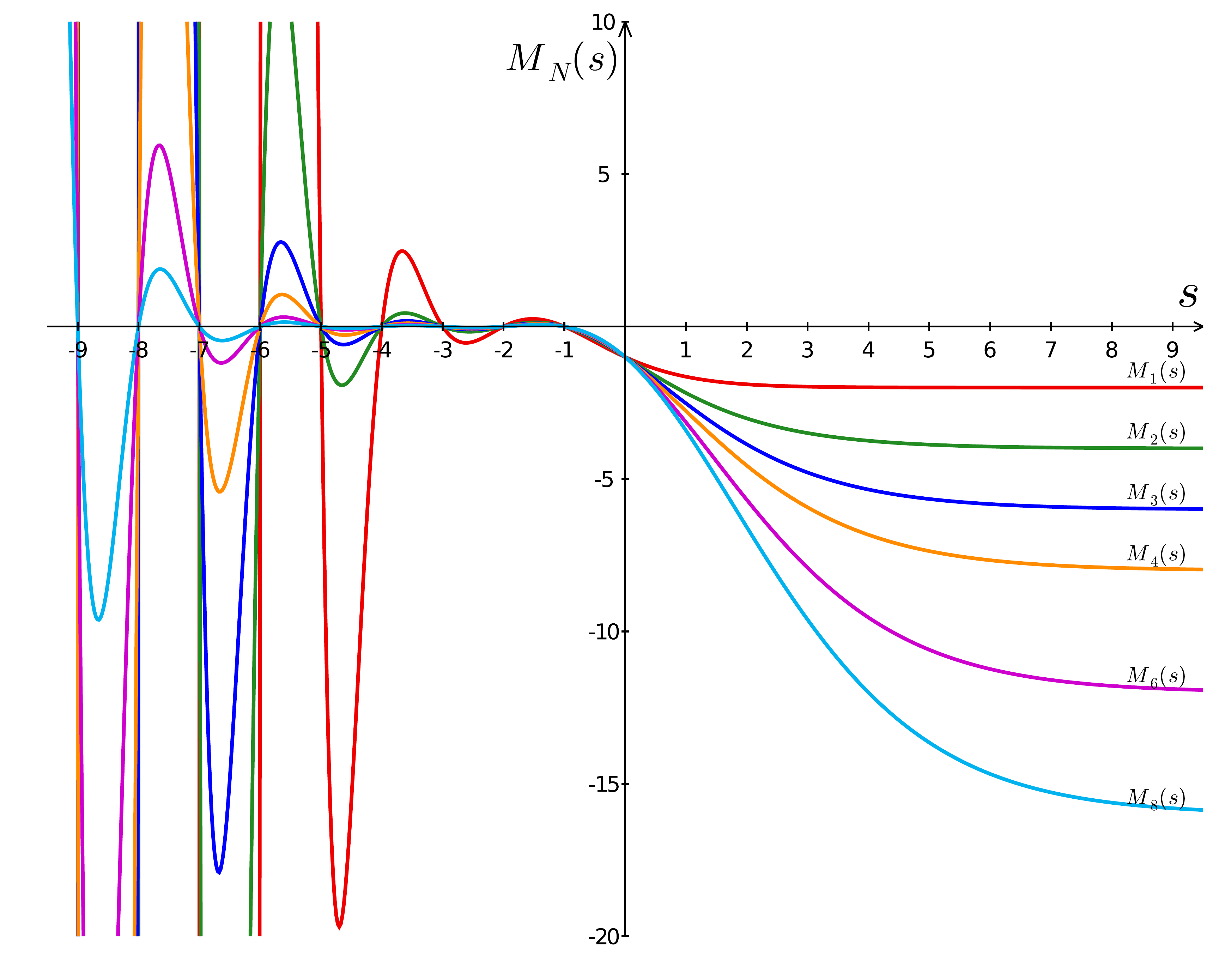}
  \caption{The Madelung constant $M_1(s), M_2(s), M_3(s), M_4(s), M_6(s)$ and $M_8(s)$ for $s\in[-9,9]$.}
  \label{fig:Madfunc1}
  \end{center}
  \end{figure}
  
Equations (\ref{Madelungrec}) and (\ref{e32}) allow for some interesting analysis. The gamma function $\Gamma(x)$ has poles at $x=0, -1, -2, \ldots$ for which the Bessel sum in (\ref{Madelungrec}) and (\ref{e32}) vanishes. In this case we get
\begin{equation}
\label{eq:limit}
M_{N}(s)  =-2\eta(2s) \quad {\rm if}~ s=0, -1, -2, \ldots 
\end{equation}
which is independent of the dimension $N$. This implies that all Madelung curves cross at these critical points. Moreover, from the Dirichlet eta function we know that $\eta(2s)=0$ for $s=-1,-2,\ldots$. This behavior is nicely seen in Figure \ref{fig:Madfunc1}. Comparing with Zucker's formulas we see that this is easily fulfilled for the specific dimensions given. Concerning the usual Madelung constant at $s=1/2$ we see that they lie close to the crossing point at $s=0$ which explains their rather slow decrease with increasing dimension $N$.

Zucker was able to evaluate the Madelung series analytically for even dimensions up to $N=8$ \cite{Zucker-1974} based on previous work of Glasser \cite{glasser1973,Glasser1973b}. He further conjectured that $M_3(s)$ may be expressed in terms of a yet unknown Dirichlet series (for a recent analysis of lattice sums arising from the Poisson equation see Ref.\cite{Bailey_2013}). Of considerable help for future investigations will be the condition that $M_N(0)=-1$ and $M_N(-n)=0$ for all $n\in\mathbb{N}$. At these critical points we have the following properties 
\begin{align}
\label{Madelungcond}
\nonumber
\zeta(0) &=  -\frac{1}{2} \quad , \quad \zeta(-2n) =  0 \quad , \quad \zeta(-n)=(-1)^n\frac{B_{n+1}}{n+1}\\
\eta(0) &=  \frac{1}{2} \quad , \quad \eta(-2n) =  0 \quad , \quad \eta(-n)=\frac{\left( 2^{n+1} -1\right)}{n+1} B_{n+1}\\
\nonumber
\beta(0) &=  \frac{1}{2} \quad , \quad \beta(-2n+1) =  0 \quad , \quad \beta(-n)=\frac{E_n}{2}
\end{align}
where $B_n$ and $E_n$ are the Bernoulli and Euler numbers respectively \cite{andrews1999special}. For example, from Zucker's formulas (\ref{MadelungN6}) and (\ref{MadelungN8}) we follow immediately that $M_6(0)=E_2=-1$ and $M_8(0)=2(2^4-1)B_4=-1$. Further,  because of  $\lim_{s\to\infty}\eta(s) = 1$, $\lim_{s\to\infty}\beta(s) = 1$, $\lim_{s\to\infty}\zeta(s) = 1$ we see that the coefficients in front of the functions in eqs.(\ref{MadelungN1})-(\ref{MadelungN8}) add up to exactly $-2N$. It is, however, incorrect to assume that analytical formulas in terms of these standard functions can be obtained for higher even dimensions. For a detailed discussion we refer to Appendix B. In this sense, our expansions in terms of Bessel functions is perhaps the closest general form for a fast convergent series we can get for the $N$-dimensional Madelung constant.

\section{Conclusions}

We presented fast convergent expressions for the Madelung constant in terms of Bessel function expansions which allow for an asymptotic exponential decay of the series. Even for higher dimensions the Madelung constants can be evaluated efficiently and accurately through the recursive expression or by using computer algebra to work with the generating functions. The number of representations of the $N$ sum of squares can also be efficiently handled through recursive relations. The Madelung constants and their analytical continuations can be calculated easily by standard mathematical software packages to any precision. These numbers may be useful for future explorations of analytical formulas in higher dimensions. For $s\ge1/2$ a Fortran program with double precision accuracy is available from our CTCP website \cite{ProgramJones}.





\appendix
\section{Special Functions} 

We give a brief overview over the special functions used in this work. More details can be found in the book by Andrews \cite{andrews1999special}. The Dirichlet (or $L$-) series (Riemann zeta, Dirichlet eta, and Dirichlet beta functions) are defined as
\begin{equation} \label{eq:RiemannZeta}
 \zeta(s) = {\sum\limits_{i \in \mathbb{N}}} i^{-s} \,,
\end{equation}
\begin{equation} \label{eq:DirichletEta}
\eta(s)=\sum\limits_{i \in \mathbb{N}} (-1)^{i-1} \, i^{-s} = \left( 1 - 2^{1-s}\right) \zeta(s) \,.
\end{equation}
\begin{equation} \label{eq:beta}
\beta(s)= \sum_{i\in \mathbb{N}}(-1)^{i+1} (2i-1)^{-s} \,.
\end{equation}
Their analytic continuations to $L$-functions into the negative real part (or the whole complex plane) are given by \cite{glasser1973}
\begin{equation} \label{eq:DirichletEtaAC}
\eta(-s)=s(2-2^{-s})\pi^{-s-1}{\rm sin}(\tfrac{\pi}{2} s)\Gamma(s)\zeta(s+1) \,.
\end{equation}
\begin{equation} \label{eq:DirichletBetaAC}
\beta(1-s)=\left(\frac{\pi}{2}\right)^{-s}{\rm sin}(\tfrac{\pi}{2} s)\Gamma(s)\beta(s)  \,.
\end{equation}
\begin{equation} \label{eq:RiemannzetaAC}
\zeta(-s)=-2^{-s}\pi^{-s-1} \left(\frac{\pi}{2}\right)^{-s}{\rm sin}(\tfrac{\pi}{2} s)\Gamma(s+1)\zeta(s+1) \,.
\end{equation}
Here, the gamma function is usually defined for real positive numbers as
\begin{equation} \label{eq:gamma}
 \Gamma(s)=\int_{\mathbb{R}_+} x^{s-1} e^{-x} dx  \,.
\end{equation}
 and when $s=n\in\mathbb{N}$ we have $\Gamma(n)=(n-1)!$. The gamma function on the whole real axis is then defined as the analytic continuation of this integral function to a meromorphic function by the simple recursion relation $\Gamma(x)=\Gamma(x+1)/x$ with $1/\Gamma(-n)=0$ for $n\in\mathbb{N}_0$  \cite{Artin2015}. 

The modified Bessel function of the second kind is defined as
\begin{equation} \label{eq:bessel4}
K_\nu (x)=\frac{1}{2} \int_{\mathbb{R}_+} u^{\nu-1}{\rm exp}\left\{ -x\left( u+u^{-1} \right)/2 \right\} du  \,,
\end{equation}
The higher-order Bessel functions can be successively reduced to lower order Bessel functions by
\begin{equation} \label{eq:Besselrecursive}
K_\nu(x)=\frac{2(\nu-1)}{x}K_{\nu-1}(x) + K_{\nu-2}(x) \,,
\end{equation}
and we use the symmetry $K_{-\nu}(x)=K_\nu(x)$ for its analytical continuation.

The representations of the sum of squares is obtained from the recursive formula
\begin{equation}
\label{recursive}
r_{N+1}(m)=r_{N}(m)+2\sum_{\substack{i\in \mathbb{N} \\ \;\; m-i^2\ge0}} r_{N}(m-i^2)
\end{equation}
keeping in mind that $r_N(0)=1$. Eq.(\ref{recursive}) can easily be derived from its generating function,
\begin{equation}
\sum_{m\in \mathbb{N}_0} r_N(m) = \left( \sum_{k\in \mathbb{Z}} q^{k^2}\right)^N  \,.
\end{equation}
In a similar fashion one obtains a recursive formula for the sum of odd squares,
\begin{equation}
\label{recursive1}
r_{N+1}^{\rm odd}(m)=2\sum_{\substack{i\in \mathbb{N} \\ m-(2i-1)^2>0}} r_{N}^{\rm odd}(m-(2i-1)^2)
\end{equation}
keeping in mind that $r_N^{\rm odd}(0)=0$ and we do not include this term in our summation. 
For completeness we mention that the sum of even squares is trivially related to the sum of squares by $r_{N}^{\rm even}(4m)=r_N(m)$ and $r_{N}^{\rm even}(m)=0$ if $m$ is not divisible by  4.

\section{Why Zucker's analytical formulas do not continue into higher dimensions}

Zucker's formulas (\ref{MadelungN1})-(\ref{MadelungN8}) are equivalent to Jacobi's formulas for sums of 2, 4, 6 and 8 squares (e.g., see \cite{Cooper-2017} pp. 177, 202, 238):
\begin{align}
\left(\sum_{j\in \mathbb{Z}} (-1)^jq^{j^2}\right)^2 &= 1-4\sum_{n\in \mathbb{N}} \chi_4(n) \frac{q^n}{1+q^n}, \label{ss2} \\
\left(\sum_{j\in \mathbb{Z}} (-1)^jq^{j^2}\right)^4 &= 1+8\sum_{j\in \mathbb{N}} \frac{j(-q)^j}{1+q^j}, \label{ss4} \\
\left(\sum_{j\in \mathbb{Z}} (-1)^jq^{j^2}\right)^6 &= 1+4\sum_{j\in \mathbb{N}} \chi_{4}(j)\frac{j^2q^j}{1+q^j} \label{ss6}
+ 16\sum_{j\in \mathbb{N}} \frac{j^2(-q)^j}{1+q^{2j}}, \\
\left(\sum_{j\in \mathbb{Z}} (-1)^jq^{j^2}\right)^8 &= 1+16\sum_{j\in \mathbb{N}} \frac{j^3(-q)^j}{1-q^j},\label{ss8}
\end{align}
respectively, where
\begin{equation}
\label{Legendre4def}
\chi_4(n) =\sin \frac{n\pi}{2} =   \begin{cases}
1 & \mbox{if $n\equiv 1\pmod{4}$}, \\
-1 & \mbox{if $n \equiv 3 \pmod{4}$}, \\
0 & \mbox{otherwise.}
\end{cases}
\end{equation}
For example, the formula~\eqref{ss8} can be written in the form
\begin{equation}
\left.\sum_{i_1,i_2,\ldots,i_8\in \mathbb{Z}}\right.^{\!\!\!\!\!\!\!\! \prime} \hspace{.3cm}(-1)^{i_1+i_2+\cdots + i_8} \; q^{i_1^2+i_2^2+\cdots + i_8^2}
=16\sum_{j\in \mathbb{N}}\sum_{k\in \mathbb{N}} j^3(-1)^j q^{jk}.
\end{equation}
Put $q=e^{-u}$, multiply both sides by $u^{s-1}$ and integrate, to obtain
\begin{equation}
\left.\sum_{i_1,i_2,\ldots,i_8\in \mathbb{Z}}\right.^{\!\!\!\!\!\!\!\! \prime} \hspace{.3cm} (-1)^{i_1+i_2+\cdots + i_8} 
\int_{\mathbb{R}_+} u^{s-1} e^{-u(i_1^2+i_2^2+\cdots + i_8^2)} \ud u
=16\sum_{j\in \mathbb{N}} \sum_{k\in \mathbb{N}} j^3(-1)^j  \int_{\mathbb{R}_+} u^{s-1}e^{-ujk} \ud u.
\end{equation}
The integrals can be evaluated using eq.(\ref{eq:gamma}) to give
\begin{align}
\left.\sum_{i_1,i_2,\ldots,i_8\in \mathbb{Z}}\right.^\prime \frac{(-1)^{i_1+i_2+\cdots + i_8} }{(i_1^2+i_2^2+\cdots + i_8^2)^s}
=16\sum_{j\in \mathbb{N}} \sum_{k\in \mathbb{N}} \frac{j^3(-1)^j}{(jk)^s},
\end{align}
where the common factor $\Gamma(s)$ has been cancelled from each side. In other words, we have obtained
\begin{align}
M_8(s) = 16\left(\sum_{j\in \mathbb{N}}  \frac{j^3(-1)^j}{j^s} \right) \left(\sum_{k\in \mathbb{N}}  \frac{1}{k^s} \right) 
= -16\left(\sum_{j\in \mathbb{N}}  \frac{(-1)^{j-1}}{j^{s-3}} \right) \left(\sum_{k\in \mathbb{N}}  \frac{1}{k^s} \right) 
= -16\eta(s-3)\zeta(s).
\end{align}
Thus we have obtained Zucker's formula (\ref{MadelungN8}) from the sum of squares formula~\eqref{ss8}.
The process is reversible, so (\ref{MadelungN8})  is equivalent to~\eqref{ss8}.
By similar calculations, each of Zucker's formulas ({\ref{MadelungN2})--({\ref{MadelungN8}) is equivalent to the
respective formula in~\eqref{ss2}--\eqref{ss8}.

By analogy with $M_8(s)$ in eq.(\ref{MadelungN8}), it is tempting to speculate that there might be expressions for $M_{10}(s)$ and $M_{12}(s)$
as finite sums of the forms
\begin{align}
M_{10}(s) = \sum_i f_i(s-4)g_i(s)  \quad {\rm and} \quad M_{12}(s) = \sum_i F_i(s-5)G_i(s) 
\end{align}
for certain $L$-functions $f_i(s)$, $g_i(s)$, $F_i(s)$ and $G_i(s)$. However this is unlikely to be true for reasons that we shall now explain.

There are formulas for sums of 10, 12, 14, $\ldots$ squares that are similar to Jacobi's~\eqref{ss2}--\eqref{ss8}, but
they involve other more complicated terms called cusp forms \cite{apostol1976}. Glaisher found the formulas for 10, 12, 14, 16 and 18 squares, and a general formula for any even number of squares was obtained by Ramanujan. The formulas for sums of 10 and 12 squares are
\begin{align}
\left(\sum_{j\in \mathbb{Z}} (-1)^jq^{j^2}\right)^{10}
&= 1 - \frac{4}{5} \sum_{j\in \mathbb{N}} \frac{\chi_4(j)j^4 q^{j}}{1+q^{j}}
+ \frac{64}{5} \sum_{j\in \mathbb{N}} \frac{j^4 (-q)^j}{1+q^{2j}}
- \frac{32}{5}E_{10}(q)
       \label{ss10}
\intertext{and}
\left(\sum_{j\in \mathbb{Z}} (-1)^jq^{j^2}\right)^{12}
&=  1 + 8 \sum_{j\in \mathbb{N}} \frac{j^5(-q)^j}{1+q^j}
    - 16E_{12}(q) 
\label{ss12}
\end{align}
where
\begin{align}
E_{10}(q) = q \prod_{j\in \mathbb{N}} \frac{(1-q^{2j})^{14}}{(1-q^j)^4}
\quad\text{and}\quad E_{12}(q) = q \prod_{j\in \mathbb{N}}(1-q^{2j})^{12}.
\end{align}
For a statement of the general formula, see \cite{Cooper-2017} (p. 214). A proof of the general formula and references to other proofs can be found in ref.\cite{cooper2001sums}.

There is no simple formula for the coefficients in the expansions of~$E_{10}(q)$ or~$E_{12}(q)$, but they satisfy some remarkable properties. For example, if we write 
\begin{equation}
E_{12}(q) = \sum_{n\in \mathbb{N}} e_{12}(n)q^n
\end{equation}
then it is known that 
\begin{equation}
e_{12}(mn) = e_{12}(m)e_{12}(n)
\end{equation}if $m$ and $n$ are relatively prime.
For prime powers, there is the three-term recurrence
\begin{equation}
e_{12}(p^{\lambda+1}) =e_{12}(p) e_{12}(p^{\lambda})  - p^5 e_{12}(p^{\lambda-1}).
\end{equation}
Furthermore, Ramanujan proved that
\begin{equation}
|e_{12}(n)| = O(n^{3+\epsilon}) \quad \text{as $n\rightarrow\infty$}
\end{equation}
and conjectured that
\begin{equation}|e_{12}(n)| \leq n^{5/2} d(n)
\end{equation}
where $d(n)$ is the number of divisors of~$n$. In fact Ramanujan had a conjecture for a sum of $2k$ squares ($k \geq 5$),
and that conjecture was proved by Deligne about 50 years later (as part of work for which he subsequently received the Fields medal).

To complete the example for the 12-dimensional lattice, if we put $q=e^{-u}$ in \eqref{ss12}, multiply by $u^{s-1}$ and integrate, the result is
\begin{equation}
\label{Z13}
M_{12}(s) = -8\eta(s-5)\eta(s) - 16\sum_{n\in \mathbb{N}} \frac{e_{12}(n)}{n^s},
\end{equation}
where the coefficients $e_{12}(n)$ are as above.
It was known to Ramanujan that the Dirichlet series can be factored, and hence we obtain the formula
\begin{equation}
\label{Z12}
M_{12}(s) = -8\eta(s-5)\eta(s) - 16\prod_{p} \frac{1}{\left(1-\frac{e_{12}(p)}{p^s}+\frac{1}{p^{2s-5}}\right)}
\end{equation}
where the product is over the odd prime values of $p$. The first few values are as follows:  $e_{12}(3)=  -12, e_{12}(5)= 54, e_{12}(7)= -88, e_{12}(11)= 540, e_{12}(13)= -418, e_{12}(17)= 594, e_{12}(19)= 836, e_{12}(23)= -4104, e_{12}(29)= -594$.

The formula~\eqref{Z12} is the analogue of Zucker's formulas for the 12-dimensional lattice.
Similar formulas can be given for sums of $2k$ squares for any positive integer~$k$.
The number of cusp forms is \mbox{$\lfloor(k-1)/4\rfloor$.} In particular, there are no cusp forms for $1\leq k \leq 4$ corresponding
to Zucker's formulas for the lattice sums in $2$, $4$, $6$ or $8$ dimensions;
there is one cusp form for $5 \leq k \leq 8$ corresponding to the lattice sums
in $10$, $12$, $14$ or $16$ dimensions; and there are two cusp forms for $9 \leq k \leq 12$ corresponding to 
the lattice sums in $18$, $20$, $22$ or $24$ dimensions.

As a consequence of Ramanujan's conjectures and Deligne's proofs, we now know that the number of representations
of $N$ as a sum of an even number $2k$ squares is given by a dominant term that involves
a sum of the $(k-1)$th powers of the divisors of~$N$, plus
a correction term (the coefficient in a cusp form) that is roughly the square root in magnitude of the dominant term.
When the number of squares is 2, 4, 6 or 8 there is no cusp form, and the divisor sum formula is exact, and that
is the reason the formulas of Zucker exist.
When the number of squares is 10, 12, 14, $\ldots$, there is an increasing number of cusp forms, and
there is no easy formula for the coefficients in their power series expansions. That is the reason why Zucker's formulas
stop at 8 dimensions, and why there are no similar formulas for dimensions 10, 12, 14, $\ldots$.

\section*{Acknowledgements} 
This work was supported by the Marsden Fund Council from Government funding, managed by the Royal Society of New Zealand (MAU1409)

\bibliographystyle{elsarticle-num} 
\bibliography{references}

\end{document}